\documentclass{aa}  
\usepackage{multirow}
\usepackage{graphicx}
\usepackage{subfig}
\usepackage{txfonts}
\begin{document}

\title{Nitrile versus isonitrile adsorption at interstellar grain surfaces II. Carbonaceous aromatic surfaces}

   \author{     M. Bertin, \inst{1}    
                   M. Doronin, \inst{1,2}
                        X. Michaut \inst{1}
                        L. Philippe \inst{1}
                   A. Markovits, \inst{2}
                J.-H. Fillion,  \inst{1} 
                        F. Pauzat, \inst{2}
                Y. Ellinger, \inst{2}    
                J.-C. Guillemin  \inst{3} 
          }

   \offprints{M. Bertin}

          \institute {LERMA, Sorbonne Universit\'{e}s, UPMC Univ. Paris 06, Observatoire de Paris, PSL Research University, CNRS, F-75252, Paris, France\\
                   \email{mathieu.bertin@upmc.fr}
         \and
                Sorbonne Universit\'{e}s, UPMC Univ. Paris 06, UMR - CNRS 7616, Laboratoire de Chimie Th\'{e}orique,  75252 Paris CEDEX 05, France\\
                \email{alexis.markovits@upmc.fr} 
        \and 
                Institut des Sciences Chimiques de Rennes, \'Ecole Nationale Sup\'erieure de Chimie de Rennes, UMR - CNRS 6226, ENSCR, F-35700 Rennes, France. \\
                \email {jean-claude.guillemin@ensc-rennes.fr}   
        }

   \date{Received ???? , 2017; accepted ???? , 2017}

  \abstract 
   { Almost 20\% of the $\sim200$ different species detected in the interstellar and circumstellar media 
   present a carbon atom linked to nitrogen by a triple bond. Of these 37 molecules, 30 are nitrile R-CN compounds, the remaining 7 belonging to the isonitrile R-NC family.  How these species behave in their interactions with the grain surfaces is still an open question. }  
   {In a previous work, we have investigated whether the difference between nitrile and isonitrile functional groups may induce differences in the adsorption energies of the related  isomers at the surfaces of interstellar grains of various nature and morphologies. This study is a follow up of this work, where we focus on the adsorption on carbonaceous aromatic surfaces. }
   {The question is addressed by means of a concerted experimental and theoretical approach of the adsorption energies of CH$_3$CN and CH$_3$NC on the surface of graphite (with and without surface defects).  The experimental determination of the molecule and surface interaction energies is carried out using temperature-programmed desorption (TPD) in an ultra-high vacuum (UHV)  between 70 and 160 K. 
     Theoretically, the question is  addressed  using first-principle periodic density functional theory (DFT) to represent  the organised solid support.}  
   {The adsorption energy of each compound is found to be very sensitive to the structural defects of the aromatic carbonaceous surface: these defects, expected to be present in a large numbers and great diversity on a realistic surface, significantly increase the average adsorption energies to more than 50\% as compared to adsorption on perfect graphene planes. The most stable isomer (CH$_3$CN) interacts more efficiently with the carbonaceous solid support than the higher energy isomer (CH$_3$NC), however.  } 
   {}

   \titlerunning{Nitrile versus isonitrile adsorption II. Aromatic carbonaceous surfaces }  
    \authorrunning{Bertin et al.}  

   \maketitle
 \section{Introduction} 
 
The C$\equiv$N nitrile bond is  the most widespread functional group among the $\sim$200 interstellar and circumstellar molecules detected to date ({\it http://www.astro.uni-koeln.de/cdms/molecules} and/or {\it http://www.astrochymist.org/astrochymist$\_$ism}). This functional group gives rise to the well-known cyanopolyynes R-CN and isocyanopolyynes R-NC series when R contains C$\equiv$C conjugated triple bonds that can spread on a terminal CN group. When R is a saturated group, the number of detected isomers is reduced to CH$_3$CN (Solomon et al 1971), CH$_3$CH$_2$CN (Johnson et al 1977), {\it n}-C$_3$H$_7$CN (Belloche et al. 2009), and {\it i}-C$_3$H$_7$CN (Belloche et al. 2014) for nitriles and only CH$_3$NC (Cernicharo et al 1988) for isonitriles. We should mention that HCN (Snyder \& Buhl 1971) and HNC (Snyder \& Buhl 1972;  Zuckerman et al 1972) are not counted at this level.

The chronology of detection shows that HCN and HNC were discovered at the same time, whereas it took 17 years to detect CH$_3$NC after CH$_3$CN was identified.  It is now established from radioastronomy data that the abundance ratio HCN/HNC varies from close to unity (Irvine \& Schloerb 1984; Schilke et al 1992; Tennekes et al 2006)  to several thousands  (Fuente et al 2003) with the object observed. It is also established that the abundance ratio CH$_3$CN/CH$_3$NC is remarkably stable with a value $\sim$50 regardless of the region in which these two molecules are simultaneously observed  (Irvine \& Schloerb 1984; Cernicharo et al 1988; Remijan et al 2005). 
These two isomers have both a linear backbone and practically the same dipole moments, $\sim$4 Debye, which cannot introduce any bias in their respective rotation spectra. Another possibility
to interpret  observed abundance ratio that has been questioned in previous works is a selective adsorption at interstellar grain surfaces (Lattelais et al. 2011; 2015).  This possibility might be decisive if one isomer could desorb while the other remained attached to the grain. 

Several types of grains have been proposed since the original
work of Greenberg (1976). Relying on mid-IR data, we selected three structurally different families. The  first two, namely, silicates (Molster et al. 2002) and  water ices  (Schutte 1999; Gibb et al. 2004; Watanabe \& Kouchi 2008; Boogert et al. 2015), have been considered in Part I of this experimental and theoretical study (Bertin et al. 2017) using a model hydroxylated quartz-$\alpha$ surface and a crystalline or amorphous water ice surface. On each of these substrates, the adsorption of the two isomers CH$_3$CN and CH$_3$NC was found to be mainly driven by hydrogen-bonding with surface dangling O--H bonds. In both cases, the adsorption energy of the single CH$_3$CN molecule was found to be more important than that of CH$_3$NC by about  30~meV. Thus, any differential adsorption and thermal desorption effect between nitrile and isonitrile is always expected to lead to a gas-phase excess of the isonitrile as compared with the condensed-phase abundance in the case of hydroxylated grain surfaces. In Part II we focus on the carbonaceous aromatic grains often modelled by polycyclic aromatic hydrocarbon (PAH) molecules of various finite sizes, whose IR bands are present in the spectra of most  objects in the interstellar medium (ISM)  (L\'eger \& Puget 1984; Allamandola, Tielens \& Barker 1985). Such aromatic carbonaceous surfaces can also be modelled using graphene planes, which can be seen as polycyclic aromatic hydrocarbon of infinite size. 

In this Part II report we follow up on Part I, using the same interdisciplinary approach, taking here highly oriented pyrolytic graphite (HOPG) as a laboratory model for graphitic grains to measure adsorption energies and  periodic density functional theory (DFT) calculations as a theoretical approach to determine adsorption energies.

\section{Experimental approach to adsorption energies} 

\subsection{Experimental method}
The experimental determination of CH$_3$CN and CH$_3$NC adsorption energies on graphite surface has been realised using the Surface Processes \& ICES (SPICES) setup of the LERMA (UPMC; Paris). The studies were realised under ultrahigh vacuum conditions (base pressure of $\sim10^{-10}$ Torr). A highly oriented polycrystalline graphite (HOPG) substrate was mounted at the cold end of a closed-cycle He cryostat, where its temperature, using a resistive heating system, can be varied from 10~K to 300~K with an absolute precision of better than 0.1~K.
 Very low partial pressures of gaseous CH$_3$CN  or freshly synthesised CH$_3$NC can be introduced in situ through a dosing tube positioned a few millimeters in front of the substrate, which is  kept at 90~K. This results in the growth of a molecular ice, whose thickness is expressed in deposited monolayers ML, with a monolayer corresponding to a saturated molecular layer onto the graphite surface ($1 ML \approx 1\times10^{15}$~molecule/cm$^{2}$). Alternatively, the amount of deposited molecules can be expressed in coverage $\theta$, defined as the ratio between deposited molecules and available adsorption sites (for a close monolayer, $\theta=1$).

The experiments were realised with commercially available CH$_3$CN (Sigma-Aldrich, 99\% purity). CH$_3$NC was  prepared using the synthesis of Schuster et al. (1973), but using trioctylamine instead of quinoline as the base. The liquid CH$_3$NC being much less stable than CH$_3$CN, it had to be kept at low temperature and protected from light. Before their introduction into the setup, several freeze-pump-thaw cycles were realised to further purify the liquids from any diluted gaseous pollutant. Finally, the purity and thermal stability of each condensed compound was checked using IR spectroscopy on the deposited pure ices (see Bertin et al. 2017 for the IR spectra and the vibrational attributions).

The temperature programmed desorption (TPD) technique was used for the adsorption energies determination, but also to calibrate the initial thicknesses of the ices. It consists of monitoring the mass signal of desorbed species by means of a quadripolar mass spectrometer (QMS) while warming up the substrate with a
constant heating rate. For both CH$_3$CN and CH$_3$NC, the signal of the mass $41$~amu was monitored, which corresponds to the intact molecule ionisation in the QMS. In a first approximation, the mass signal $I(T)$, which is proportional to the desorption flux $\Phi_{des}(T)$, follows the Polanyi-Wigner law (Redhead 1962),
\begin{equation}
I(T) \propto \Phi_{des}(T) = -\frac{d\theta(T)}{dT}=\frac{\nu}{\beta}\theta^n(T)\textrm{exp}\left(-\frac{E_{ads}}{kT}\right)
,\end{equation}
where $\beta$ is the heating rate, $n$ the kinetic order of the desorption, $k$ the Boltzmann constant, $E_{ads}$ the adsorption energy, and $\nu$ a pre-exponential factor. In the transition state theory, this prefactor is related to the ratio between the partition functions of the molecule in the adsorption well and in the gas, and it can vary from species to species by several orders of magnitude (M\"uller et al. 2003). The kinetic order $n$ is usually close to 0 in the case of multilayers (i.e. ices thicker than 1~ML), whereas it is larger, and usually close to 1, in the case of the desorption of a single physisorbed monolayer and less. This transition from a 0 to a higher order kinetics can be identified from a series of TPD curves of ices with different initial coverage. The critical coverage for which the transition takes place therefore corresponds to one monolayer. This method allows for estimating the initial thickness of our ices with a relative precision in the order of 10~\% (Doronin et al. 2015, Bertin et al. 2017). The integral of the TPD curve being proportional to the total amount of initially adsorbed molecules, the desorption flux $\Phi_{des}(T)$ can be obtained by dividing the mass signal $I(T)$ by the integral of the 1~ML TPD curve. From this calibrated signal, the prefactors and adsorption energies can be derived
as a function of the sample temperature. 

\subsection{Adsorption energies on graphite}

The first part of this study (Bertin et al. 2017) focused on the thermal desorption of CH$_3$CN and CH$_3$NC from pure thick ices and from hydroxylated substrates (quartz and water-ice surfaces). Here, we apply the same experimental protocol to determine the adsorption energies of the two isomers on the graphite surface, which is explained in details in Bertin et al. 2017 and Doronin et al. 2015. A series of TPD from thin ices (submonolayer coverages) of CH$_3$CN and CH$_3$NC deposited on HOPG at 90~K was performed. The resulting curves, obtained using a heating ramp of  12~K/min, are shown in figure 1. 

\begin{figure}
\includegraphics
[width=8cm]{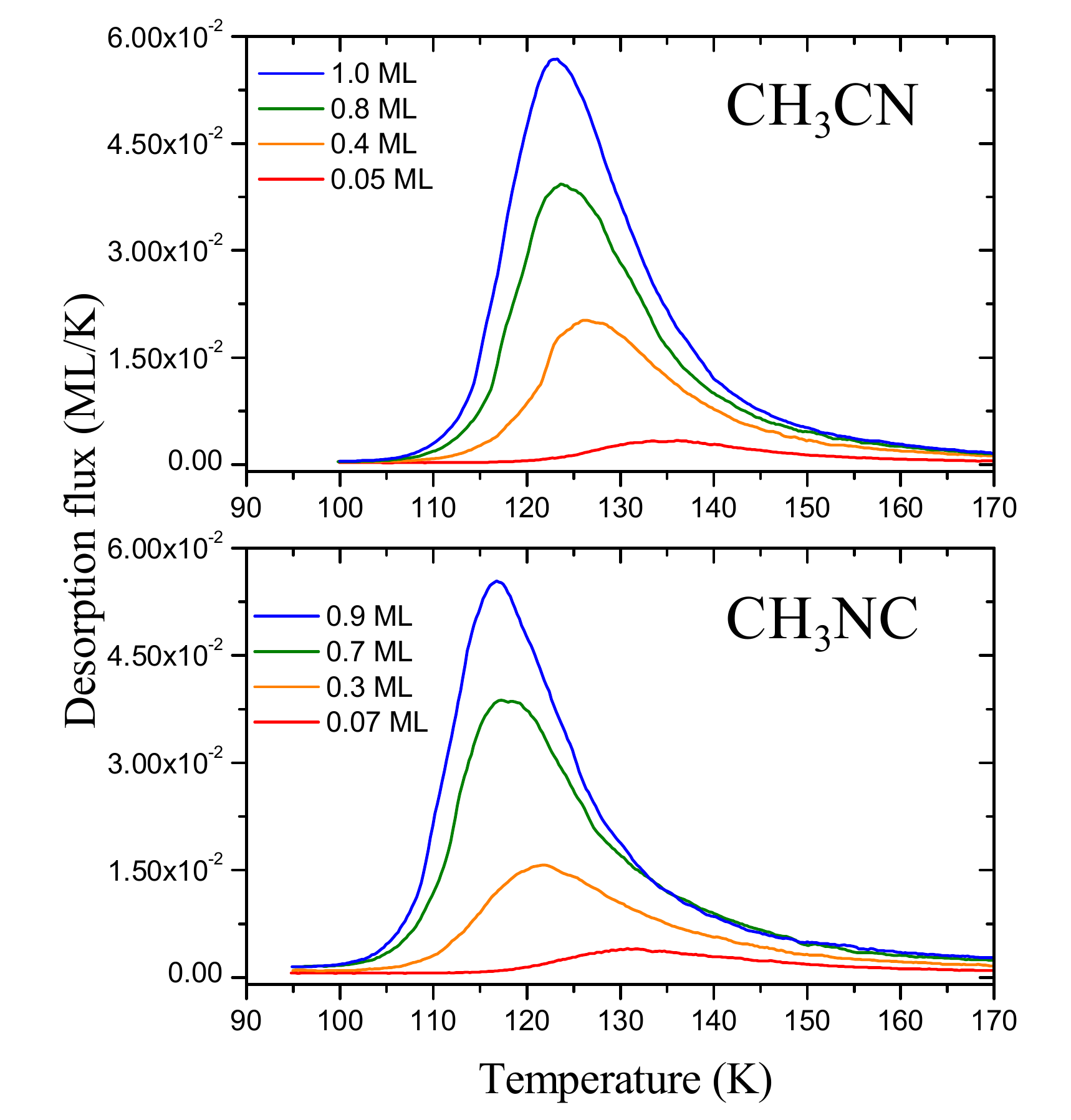} 
\caption{TPD curves obtained from submonolayer coverages of CH$_3$CN (top) and CH$_3$NC (bottom), deposited on HOPG at 90~K. The heating rate used was 12~K/min. }
\end{figure}

To extract the adsorption energies, each of these curves was fitted using the Polanyi-Wigner equation. To limit the amount of free parameters, we make the simple assumption in our model that the desorption of a monolayer and submonolayer follows a first-order kinetics. In practice this implies that the effects of self-organisation of the first layer on the desorption dynamics, such as 2D island structures, are neglected. However, the method takes into account that all the molecules on the surface are not equivalent by considering a distribution of adsorption energies instead of a single value. The TPD curves of fig. 1 were fitted using the following equation:
\begin{equation}
\Phi_{des}(T)=\frac{\nu}{\beta}\sum_i\theta_i(T)\exp\left({-\frac{E_{i}}{kT}}\right)
,\end{equation}
where $\theta_i$ is the coverage of the molecules bound to the surface with the adsorption energy $E_i$. For the energies $E_i$, we chose a constant sampling with steps of 10~meV in the 350 - 600~meV range. The fitting procedure gives as a result the initial coverages associated with each adsorption energy sample, that is, the adsorption energy distribution of the molecules on the surface. The adsorption energy distributions for CH$_3$CN and CH$_3$NC on graphite are shown in figure 2. The prefactor we employed was determined by performing the TPD of identical ices, but with different heating rates. The good value of $\nu$ was chosen as the value for which the fitting procedure gives the same energy distribution regardless of the applied heating rate. For the desorption of submonolayers of CH$_3$CN and CH$_3$NC on graphite, the prefactor was determined as $8\times10^{17\pm0.5}~s^{-1}$ and $2\times10^{16\pm0.5}~s^{-1}$ , respectively.

\begin{figure}
\includegraphics
[width=9cm]{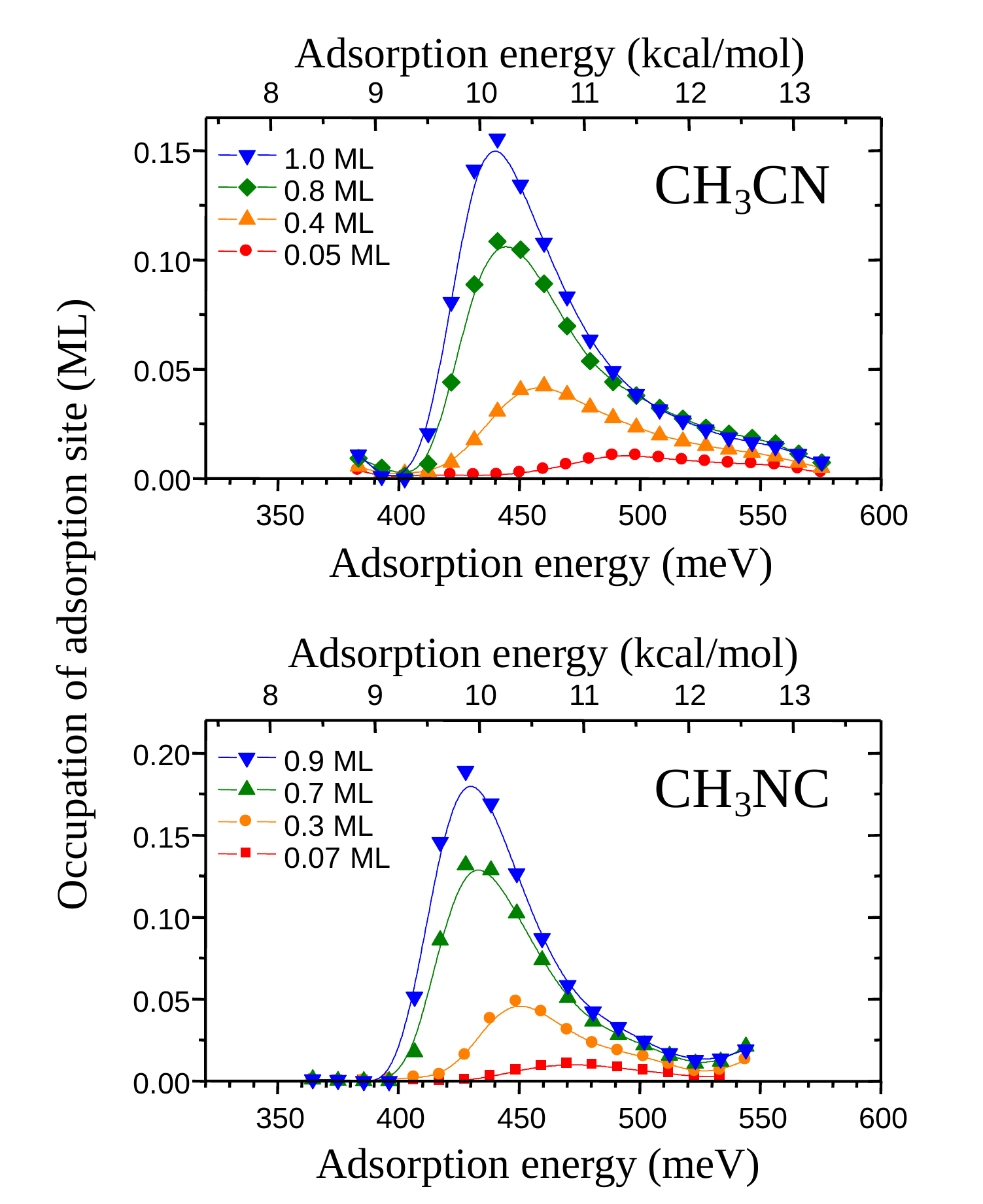} 
\caption{Adsorption energy distributions for several initial submonolayer coverages of CH$_3$CN (top) and CH$_3$NC (bottom) adsorbed on HOPG.}
\end{figure}

For each isomer adsorption on graphite, we can extract an average adsorption energy $E_{ads}$, taken as the maximum of the energy distribution, together with the width of the distribution $\delta E_{ads,}$ taken as the full width at half-maximum (FWHM) of the distribution. This width gives information on the number and dispersion of different adsorption sites on the graphite surface, but also on the deviation to the first-order kinetics approximation due to the 2D or 3D cluster-like organisation of the first molecular layer. The values we obtain for the CH$_3$CN and CH$_3$NC adsorption on graphite are presented in Table 1.

\begin{table*}
\caption{Experimental values for the pre-exponential factors $\nu$, mean adsorption energies $E_{ads}$ , and size of the adsorption energy distribution $\delta E_{ads}$ of submonolayers of CH$_3$CN and CH$_3$NC on HOPG for two initial coverages.}
\centering 
\begin{tabular}{cccccc}
\hline\hline 
 & \multirow{2}{*}{Prefactor $\nu$ (s$^{-1}$)} & \multicolumn{2}{c}{E$_{ads}$ (meV)} & \multicolumn{2}{c}{$\delta$E$_{ads}$ (meV)} \\
& & 0.7 - 1 ML &  0.3 ML & 0.7 - 1 ML & 0.3 ML \\
 \hline
\\
 CH$_3$CN on HOPG & $8\times 10^{17\pm0.5}$ & 440 & 460 & 50 & 65\\
 CH$_3$NC on HOPG & $2\times 10^{16\pm0.5}$ & 430 & 450 & 50 & 80 \\
\hline
\end{tabular} 
\end{table*}
Figure 2 shows that both CH$_3$CN and CH$_3$NC mean adsorption energies on graphite are shifted towards higher energy with decreasing initial coverage. This is expected, since in the case of smaller coverages, the molecules are free to access the most tightly bounded adsorption site during the deposition, or during the warming up, before their desorption. For higher coverages, these sites being already occupied, the excess molecules will probe less tightly bounded sites, which will have the effect of lowering the measured mean adsorption energy. However, these shifts are relatively small: in the 1 - 0.3 ML coverage range, the mean adsorption energy is shifted by only 20 meV for both isomers.  For smaller initial coverages (less than 0.1 ML), this shift seems to be stronger, but the low signal-to-noise ratio in this case also gives much higher uncertainties on the mean adsorption energy values.

By comparing the adsorption energies of each isomer, we find average adsorption energies of CH$_3$CN slightly higher than that of CH$_3$NC (of about 10 meV). However, this difference is small as compared with the size of the adsorption energy distribution, and falls within the experimental uncertainties.

 \section{Theoretical approach to adsorption energies} 
 
 The adsorption energy  E$_{ads}$    is the binding energy of a chemical species, atom, or molecule (adsorbate), with a surface (substrate) on which it resides.
 
 This interaction energy   is obtained  as 
  \begin{equation}
    E_{ads} = (E_{surf} + E_{mol}) - E_{surf+mol}
    ,\end{equation}
 
 \noindent where E$_{mol}$ is here the energy of a single CH$_3$CN or CH$_3$NC molecule, E$_{surf}$ the energy of the pristine surface of the substrate, and E$_{surf+mol}$ is the  total energy of the [surface +  CH$_3$CN or CH$_3$NC] complex; all entities are optimised in isolation. 
 
 In Part I (Bertin et al. 2017), we focused on the adsorption on hydroxylated solids, namely, water ice and silica, whose surfaces present a large distribution of electronegative oxygen sites and dangling hydrogen bonds.  There the interaction between the surface and the adsorbate can be seen as  a local mechanism in view of the well-localised electronic structure of the substrate. Here in Part II, we have an opposite situation since the graphite substrate shows a delocalised density. 
  
 Several models have been proposed to represent the top layer of carbonaceous aromatic grains on which the adsorption takes place. 
 A first attempt considered polycyclic aromatic hydrocarbon (PAHs) substrates of increasing but limited dimensions 
 (Tran et al. 2002; Heine et al. 2004; Ferre-Vilaplana 2005). 
 The next model in size is graphene, that is, an infinite planar sheet formed of hexagonal carbon rings that can also be viewed as a polycyclic aromatic hydrocarbon  of infinite dimensions 
 (Arellano et al. 2000; Pauzat et al. 2011).
 Here we chose a real graphite structure, that is, a 3D modelling approach, which takes advantage of the specificity of periodic calculations  capable of replicating the unit cell along the three directions of space. 
  The sizes of the unit cell,   whose dimensions are critical parameters,  were determined so as to avoid any spurious lateral and vertical interactions.

 In the case of atomic adsorption, three plausible  sites  may be considered according to the surface topology, namely, on top of 
 
 \indent {\bf $\bullet$}  a carbon atom \\
 \indent {$\bullet$}   the centre of a benzene ring \\
 \indent {\bf $\bullet$}  the middle of a CC bond. 
 
 \noindent   What looks simple for atoms becomes more complicated for complex molecules such as CH$_3$CN and CH$_3$NC. Three different orientations can be distinguished for each adsorption site according to the orientations of the linear backbone of the heavy atoms and the position of the  CH$_3$ groups, whose rotation yields additional  degrees of freedom. 
   All these possibilities were systemically investigated. In particular, we verified that the CCN/CNC backbone remains linear in all situations (no deviation  greater than 1 degree was detected).  
    The final results were obtained in an even larger unit cell (3$\times$5$\times$2) after convergence of both the energy and the forces acting on the atoms of the system,   that is, a cell of   (12.82$\times$12.32$\times$18)$\AA^3$.

    \subsection{Adsorption energies on perfect graphite surface} 
    
    The following study focuses on the graphite ideal surface
alone.    
    
{\it i)} 
For a vertical disposition of  CH$_3$NC and CH$_3$CN with the methyl umbrella away from the surface, the  adsorption energy is 92 and 108 meV on average  for  CH$_3$NC and CH$_3$CN, respectively. The strongest adsorption occurs in both cases when the CCN/CNC  backbones are on top of a carbon atom. 

{\it ii)} In the reverse position, with the methyl umbrella pointing to the surface, 
  the  adsorption energy is   158 and 162 meV on average  for  CH$_3$NC and CH$_3$CN, respectively, except for the position above the middle of a CC bond, which is less favourable for CH$_3$NC. No difference can be seen between  CH$_3$NC and CH$_3$CN or between the various adsorption sites. The adsorption energy is stronger when the CH$_3$ hydrogens are close to the surface because of the polarisation of the CH bonds by the CN and NC groups. 

{\it iii)}  For parallel orientations of  CH$_3$NC and CH$_3$CN with respect to the graphite surface, 
a manifold of 
 24 different geometries was probed (counting rotations of the CH$_3$ group). For each isomer, all adsorption energies were gathered within  a range of $\sim$30 meV, giving  average values of 255  and 275 meV for  CH$_3$NC and CH$_3$CN, respectively. In both isomers we found a slight preference for geometries in which the CH$_3$ hydrogens are closer to the CC nuclear frame. 
 
Although the difference between the adsorption energies of the   nitrile and isonitrile isomers is small,  the same trend prevails for both isomers in all classes, namely, E$_{ads}$(CH$_3$CN) $>$ E$_{ads}$(CH$_3$NC), as in the experiments. At this point it should be stressed that the equilibrium structures found for vertical positions ({\it i}  and {\it ii})  are metastable minima that fall onto the strongest adsorption sites ({\it iii}) for any small displacement of the CCN/CNC backbone out of orthogonality with respect to the graphite top layer. Consequently, we focus on the parallel situations.

 \subsection{Adsorption energies on graphite surface with structural defects }

\begin{figure*}
\centering
\includegraphics
[width=16cm]{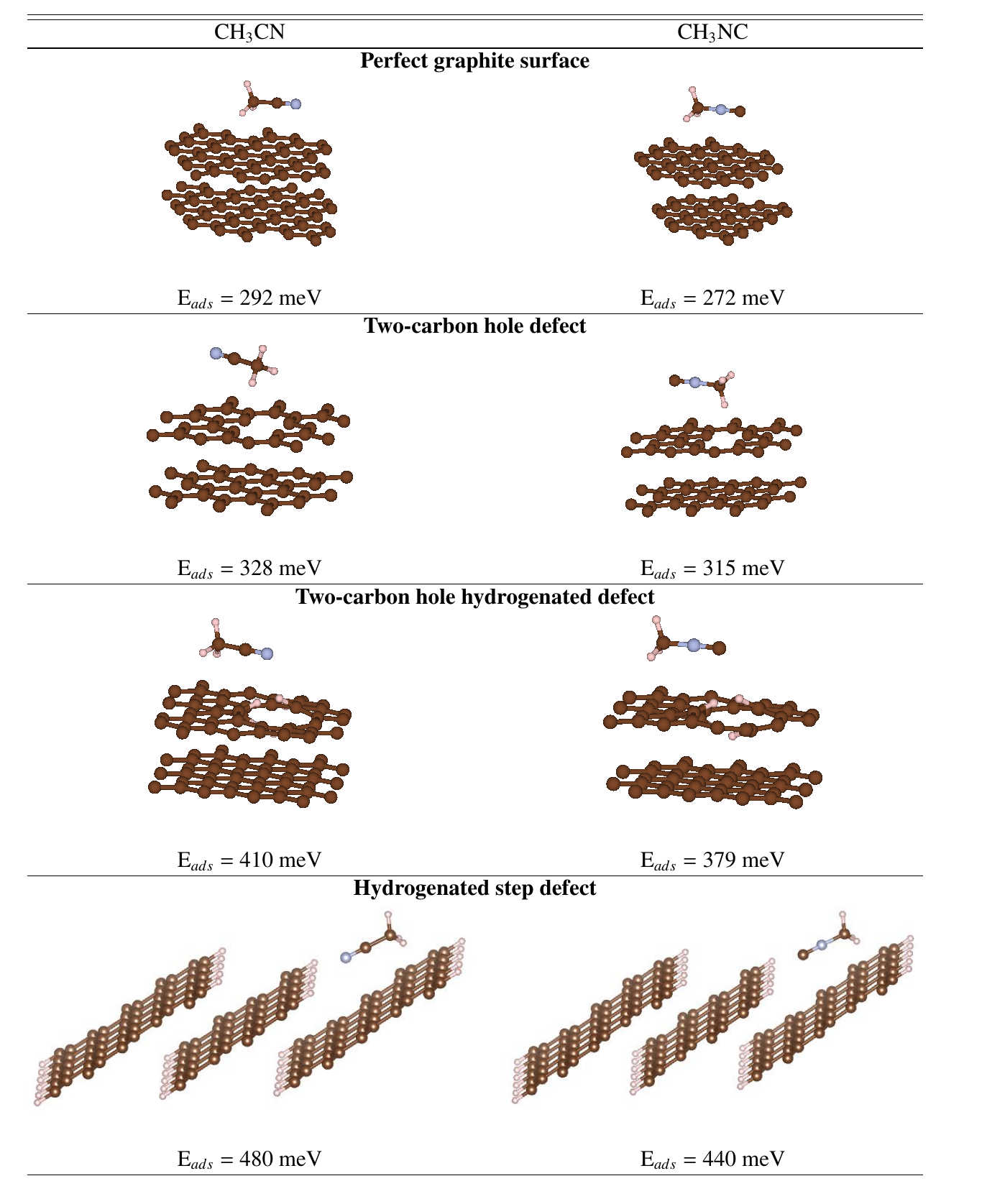} 
\caption{Views of the most stable geometries of the adsorbed CH$_3$CN and CH$_3$NC isomers on graphite.}
\end{figure*}

\smallskip
   
 It is well understood that there is no perfect surface in space nor in the laboratory, and that the range of possible defects is too wide to be modelled exhaustively. Nonetheless, to better
match the model surface to the experimental surface, several types of imperfections were  {\it \textup{a priori}} introduced in the topmost layer. In order to make it clear in the discussion, we present in Figure 3 as examples  the most stable adsorption sites on the different graphite surfaces we considered.

{\it i)} 
With the removal of two adjacent carbon atoms, leaving a localised hole the size of pyrene (Figure 3), the CH$_3$NC and CH$_3$CN isomers are no longer strictly parallel to the surface. The methyl groups are slightly tilted towards the hole to favour the interaction of the CH bonds with the electrons released on the edge of the hole. The adsorption energies are then increased by $\sim10\%$.

{\it ii)} 
A related type of  localised defect was also considered by adding supplementary hydrogen atoms to the four carbons at the edge of the hole (Figure 3), which caused the top layer to locally resemble a hydrogenated amorphous carbon (Papoular et al 1995; Colangeli et al 1995, Pauzat \& Ellinger, 2001). In this case, the CN/NC bonds are above the edges of the hole, interacting with the added hydrogens.  The adsorption energies are then increased by  $\sim30\%$ with respect to the pristine surface.

{\it iii)} 
 Another type of defect, delocalised on the surface, is provided by steps. In this case, part of the topmost layer has been  pulled up,  leaving a terrace placed on top of the layer underneath. This is typical of the making of graphite samples in the laboratory. 
It should be specified that the steps are described using the ({10$\bar{1}$5}) surface (more details can be found in Buono et al. 2014). In view of the ubiquitous presence of hydrogen,  we only considered the hydrogenated  step. Here, both isomers show a double interaction, implying the two extremities of each molecule, with the two terraces defining the step. 
One is the interaction of N for CH$_3$CN (C for CH$_3$NC, respectively) with a hydrogen sticking out of the edge of the upper terrace, the other is the interaction of the hydrogens of the CH$_3$ group with the aromatic plane of the lower terrace. The addition of these two effects is at the origin of the drastic increase in adsorption energies  by $\sim50\%$.

In all cases, the nitrile is more tightly bound to the surface than the isonitrile, regardless of the nature of the surface, that is, a perfect or a damaged polyaromatic sheet.

   \section{Discussion and final remarks}

\begin{table*}
\caption{Summary of the adsorption energies of CH$_3$CN and CH$_3$NC on graphite surfaces (this work) and on $\alpha$-quartz and crystalline water-ice surfaces (Bertin et al. 2017) determined by calculations and experiments. The displayed experimental values are extracted from the 0.7-1 ML thick ices adsorbed on HOPG, quartz, and water-ice surfaces. Uncertainties are given as half of the adsorption energy distribution size. The theoretical values presented are obtained by averaging  the adsorption energies of all relevant adsorption sites over the  distribution spread.} 
\centering 
\begin{tabular}{ccccccccc}
\hline\hline 
 & & \multicolumn{2}{c}{Theoretical E$_{ads}$ (meV)}  & \multicolumn{2}{c}{Experimental E$_{ads}$ (meV)} \\
 & & CH$_3$CN  (spread)& CH$_3$NC   (spread) & CH$_3$CN & CH$_3$NC \\
\hline
\multirow{4}{*}{Carbonaceous surfaces} & Perfect graphite surface & 275  (30)& 255  (30)& \multirow{4}{*}{440 $\pm$ 25} & \multirow{4}{*}{430 $\pm$ 25} \\
& Two-carbon hole defect & 310  (40)& 300  (40)&  & \\
& Two-carbon hole hydrogenated defect & 390   (50)& 350   (50)&  & \\
& Hydrogenated step defect & 480   (0)& 440   (0)&  & \\
\hline
\multirow{2}{*}{Hydroxylated surfaces} & $\alpha$-quartz (0001) & 460 & 414 & 460 $\pm$ 30 & 430 $\pm$ 25 \\
& A-polar crystalline ice $Ih$ & 558 & 545 & * 565 $\pm$ 25 & * 540 $\pm$ 15 \\
\hline
\end{tabular} 
\tablefoot{Experimental values marked with an asterisk  should be considered with caution since the sublimation of the supporting water-ice layer plays an important role in the observed experimental desorption features.}
\end{table*}  

We have applied a joint experimental and theoretical approach to determine the adsorption energies of the isomers CH$_3$CN and CH$_3$NC onto graphite surfaces. This approach has proven to be powerful in the case of adsorption of these two isomers on hydroxylated surfaces, that is, on quartz and water-ice surfaces (see Part I of this study; Bertin et al. 2017), for which  the experimental results compared very well to the calculated adsorption energies on perfect model surfaces. Table 2 shows a summary of these results, together with the results obtained on the graphite surfaces.

Experimentally, average adsorption energies of CH$_3$CN and CH$_3$NC onto graphite are found to be relatively similar,  CH$_3$CN being slightly more tightly bound to the carbon surface than CH$_3$NC, and varying in the 430 - 460 meV range, depending on the initial adsorbate coverage. Calculations performed on perfect graphene planes result in much lower adsorption energy values: even the highest calculated adsorption energies are indeed found at 272 meV and 292 meV for CH$_3$NC  and CH$_3$CN, respectively. This gap between experimental and calculated values may be explained by the important role played by structural defects on the graphite surface. As shown by the numerical simulations, adsorption energies are very sensitive to imperfections of the graphite surface. Very simple model defects, such as two-atom holes, hydrogenated or not, lead to a strong increase in the adsorption energies of both isomers that can reach 100 meV. Other types of defects, such as hydrogenated step-edges between graphene planes, represent even more stable adsorption sites, with adsorption energies reaching 440 and 480 meV for CH$_3$NC and CH$_3$CN, respectively. The resulting adsorption energies then fall into what is experimentally measured on a real graphite substrate. Previous studies of methanol adsorption on the graphite surface used in the experiment (Doronin et al. 2015) at the time also suggested many defects on our substrate. We therefore expect these defects to be responsible for the observed high experimental adsorption energies of CH$_3$CN and CH$_3$NC on graphite as compared with what is calculated on perfect graphene planes.

This highlights that the adsorption of the nitrile and isonitrile on aromatic carbonaceous surface is strongly influenced by the defects and is not only due to the perfectly organised grid of aromatic carbon atoms in the graphene planes.  It suggests that the structural defects drive the adsorption of nitrile and isonitrile on "realistic" carbonaceous surfaces, as is here illustrated by the great proximity of experimental values to the value calculated on model defects. From an astrophysical point of view, it also suggests that when we study adsorption energies on carbonaceous surface of grains, the use of energies determined from perfectly organised surfaces may lead to a strong underestimation of the real adsorption energy. 

Interestingly, the defects play a critical role in explaining
the discrepancies between the experimental and theoretical values of adsorption energies of CH$_3$NC and CH$_3$CN on graphite, whereas in the case of hydroxylated surfaces (i.e. water-ice and quartz surface), calculations performed on perfect crystalline surfaces compared very well with the experiments (Bertin et al. 2017 and Table 2). For instance, the adsorption energy values calculated on crystalline water surfaces gave remarkably similar
results to what was experimentally derived for amorphous disorganised compact water-ice surfaces. This shows that in the cases of hydroxylated surfaces, the structural defects in the crystalline organisation play a minor role in the determination of the adsorption energy. In the latter cases, the adsorption is indeed driven by the hydrogen bonding between the nitrile or isonitrile with the dangling O--H bonds, resulting in relatively strong adsorption energies. Defects are therefore not expected to significantly change the interactions since the dangling O--H are already present on the perfectly organised surfaces. In contrast, on the perfect graphite surfaces, only purely Van der Waals interactions are responsible for the adsorption, which are quite low compared to the strength of hydrogen bonding, and therefore dangling bounds due to defects have a strong effect on the adsorption energies.

Finally, the results summarised in Table 2 show that the adsorption energies of both nitrile and isonitrile are similar for quartz and defective graphite supporting surfaces, whereas they are much higher when the molecules are condensed onto the water-ice surface. In the latter case, the adsorption energy is estimated to be comparable to the cohesion energy of the supporting water-ice itself. However, similar trends are found when we compare the difference in adsorption energies between each isomer. In the case of carbonaceous surfaces, theoretical results show that the adsorption energy of CH$_3$CN is higher than that of CH$_3$NC in any case (perfect graphene, surface, holes and hydrogenated holes, and hydrogenated step-edges) by around 10-30 meV. This is in agreement with the experiments, which also suggest a higher average adsorption energy on graphite for CH$_3$CN than for CH$_3$NC of about 10 meV, even though this energetic difference is small compared with the energetic spread of the distribution. Interestingly, this trend has also been observed in the case of the nitrile and isonitrile adsorption onto the model hydroxylated surfaces. Table 2 shows that the adsorption energy of CH$_3$CN indeed exceeds that of CH$_3$NC by about 30~meV both on quartz and water-ice surfaces. Thus, if the adsorption-thermal desorption from grains conditions the gas-phase abundance ratio of these two isomers in the ISM, any differential desorption effect should be the same regardless of the nature of the support surface. This allows for a general prediction that the nitrile isomer should be more tightly bound to the grain surface than the isonitrile by a few tens of meV, thus leading to a gas-phase enrichment of the isonitrile at a given grain temperature compared with the abundance ratio of the two isomers in the condensed phase. This finding does not depend on the nature of the grain surface. Therefore, we would expect this enrichment due to differential adsorption-thermal desorption to be the same in colder regions, where the grains are covered with H$_2$O-rich icy mantles, and in warmer regions, where the naked grain surfaces, expected to be silicated, carbonaceous or a mixture of both, could also participate in the adsorption-desorption cycles of CH$_3$CN and CH$_3$NC.

\begin{acknowledgements}
  This work was supported by the Programme National "Physique et Chimie du Milieu Interstellaire" (PCMI) of CNRS/INSU with INC/INP co-funded by CEA and CNES, and the COST Action CM 1401, "Our astrochemical history". M.D. acknowledges PhD funding from the LabEx MiChem, part of  the French state funds managed by the ANR within the investissements d'avenir program under reference ANR-11-10EX-0004-02.

  \end{acknowledgements}


\end{document}